\documentclass[showkeys,nofootinbib,
 jor,
 amsmath,amssymb,
]{revtex4-2}
\usepackage{graphicx}
\usepackage{dcolumn}
\usepackage{bm}
\usepackage{xcolor}
\begin{document}


\title{Probing the Topology of the Early Universe using CMB Temperature and Polarization Anisotropies} 
\author{Miguel-Angel Sanchis-Lozano}
 \email{miguel.angel.sanchis@ific.uv.es}
\affiliation{Instituto de Física Corpuscular (IFIC), CSIC-University of Valencia, 46980 Paterna, Spain
}


\begin{abstract}
The temperature and polarization anisotropies of the cosmic microwave background (CMB), as measured today, offer key insights into the topology of the early universe prior to inflation, for example, by discriminating between flat and warped geometries. In this paper, we focus on a Kaluza–Klein model with an extra spatial dimension that compactifies at the Grand Unified Theory (GUT) epoch, subject to mixed Neumann/Dirichlet boundary conditions at fixed points. As a consequence, a set of infrared cutoffs emerges in both the scalar and tensor spectra, leading to observable consequences in the CMB. We examine in detail the possible signatures of such a topology, particularly in relation to the even–odd parity imbalance already reported by the COBE, WMAP and {\it Planck} missions in the temperature angular correlations.
Furthermore, we extend our analysis to the existing {\it Planck} E-mode polarization data, and to the high-precision B-mode polarization measurements expected from the forthcoming {\it LiteBIRD} mission.
\end{abstract}


\maketitle

\section{Introduction}

The inflationary scenario was originally proposed in cosmology to resolve several fundamental inconsistencies of the standard Big Bang theory, which assumed that the universe began expanding at a comparatively modest rate. As a result, problems such as the horizon, flatness, and monopole problems \cite{Kolb(1994)} remained unsolved within the framework of the original "hot Big Bang" model until the advent of the inflationary paradigm \cite{Guth:1981,Linde:1982}.

Focusing on the horizon problem \cite{Ryden:2017}, causality-based arguments highlight the difficulty of explaining the observed isotropy and homogeneity of the universe on large scales, including the minute anisotropies in the Cosmic Microwave Background (CMB), without invoking a period of rapid exponential expansion. Inflation proposes that the early universe underwent such a phase, expanding far beyond the observable Hubble radius within a tiny fraction of a second \cite{Baumann:2009ds}. Once inflation ended, the physical Hubble radius began to grow, gradually encompassing regions that define the observable universe over cosmic time. The mechanism driving the exponential expansion is typically modeled (for simplicity) by a single slowly rolling scalar field known as the {\it inflaton}, whose fundamental nature remains unknown.

According to the inflationary paradigm, both primordial scalar (matter density) quantum perturbations and tensor (metric) perturbations generated temperature anisotropies superimposed on an otherwise isotropic and homogeneous background. These fluctuations were enormously amplified during the inflationary phase, making them observable today, for instance in the CMB and large scale structures. However, tensor modes contribute to CMB much less than scalar modes, which makes it difficult to disentangle their respective contributions from temperature anisotropies alone. In contrast, primordial gravitational waves (PGWs) are expected to generate B-mode polarization in the CMB, whereas scalar fluctuations do not (to first order in perturbation theory) \cite{Kamionkowski:2015yta}. In fact, the detection of PGWs is widely regarded as one of the key objectives of inflationary cosmology.

However, sources other than PGWs can also induce B-mode polarization, such as gravitational lensing or foreground contamination of Galactic origin, and their contributions must be carefully separated from those of PGWs before any discovery claim can be made. Let us point out that the expected pattern generated by PGWs in the CMB power spectrum peaks both at small angular scales (corresponding to multipoles of order $\ell \simeq 80$) and at large angular scales (low multipoles, of order $\ell \lesssim 10$). In this paper, we focus on the latter, which, besides its relevance for PGWs, could also provide information on the topology of the early universe, this constituting the main goal of this paper.

A standard approach to studying PGWs is to parametrize the tensor power spectrum in terms of the tensor-to-scalar ratio, $r$, and the spectral index, $n_s$, and to constrain them using measurements of the temperature and polarization spectra. In this paper, following previous work \cite{sanchis-lozano:2022s,sanchis-sanz:2024}, we analyze the temperature, E- and B-mode polarization angular correlations (with special emphasis on the even–odd (im)balance of multipoles) which could potentially reveal or constrain the topological structure of the early universe \cite{Brandenberger:1994,Planck:2015gmu}.

Admittedly, the current modest significance of the odd-parity preference, among other anomalies deviating from the Standard Cosmological Model observed in the CMB by all three satellite missions—COBE \cite{COBE:1996}, WMAP \cite{WMAP:2003}, and {\it Planck} \cite{Planck:2014,Planck:2018jri,Planck:2019evm}, could simply be due to a statistical fluctuation. On the other hand, a possible connection between the odd-parity preference and the lack of large-angle correlations has been contemplated in the literature, though no clear theoretical explanation has yet emerged \cite{Land(2005),Kim(2012),Schwarz(2016),Copi(2018)}. However, if both effects were to have a (common or not) physical origin, their implications would be extremely important, and further investigation is certainly worthwhile. 

To this end, we first review the CMB temperature anisotropies by revisiting our previous work and extending it to include polarization, while further developing the specific theoretical framework relevant in our approach \cite{sanchis-sanz:2024}. On the other hand, recent studies 
\cite{Liu:2024mvp,Liu-Melia:2025} provide a comprehensive analysis of CMB temperature and E-mode and B-mode polarization anisotropies, introducing a single infrared (IR) cutoff in the scalar and tensor spectra and comparing it with the case without a cutoff. 
In fact, the possibility of introducing an IR cutoff in the scalar power spectrum has been considered in quite many occasions in the literature attributed to different origins \cite{Contaldi:2003zv,Cline:2003ve,Kuhnel:2010pp,Hazra:2014jwa,Gruppuso:2016,Ashtekar:2021izi}). In the present work, we extend this scenario by introducing two IR cutoffs, rather than a single one, in both the scalar and tensor spectra, and perform a comparative analysis across all cases.

 Finally, we examine future precision measurement for B-mode polarization through angular correlations by the {\it LiteBIRD} mission \cite{LiteBIRD:2022cnt}. Such measurements will certainly shed light on the quantum nature of gravity and uncover potential new physics beyond the Standard Model, as CMB photons are expected to carry information from epochs preceding inflation. These signals, sensible to tensor modes, will likely complement, or even surpass, the information derived from temperature angular correlations, especially regarding the topology of the universe, which is the central focus of this work.

\section{Temperature anisotropies}

One of the most outstanding results from the CMB measurements is its extraordinary homogeneity across the full sky, reflecting the homogeneity of the initial source of perturbations itself, extending over the whole visible universe today. This fact, in turn, implies isotropy along all directions of sky. 

However, small but significant departures from this behavior contain important information about the early stages of the universe for both scalar and tensor modes contributing to the CMB. Still, correlations between temperature anisotropies at different points in the sky depend only on the angle between them. Then, the two-point angular correlation function is the appropriate tool for analyzing these fluctuations. It is defined as the ensemble product
of the CMB temperature differences $\delta T(\vec{n})$ with respect to the average temperature from all pairs of directions
in the sky defined by unitary vectors $\vec{n}_1$ and $\vec{n}_2$:
\begin{equation}\label{eq:CTT}
C^{\rm TT}(\theta)=\langle \delta T(\vec{n}_1)\ \delta T(\vec{n}_2) \rangle\;,
\end{equation}
where $\theta \in [0,\pi]$ is the angle defined by the scalar product $\vec{n}_1 \cdot \vec{n}_2$.

\begin{figure*}

\centering
\includegraphics[width=9.6cm]{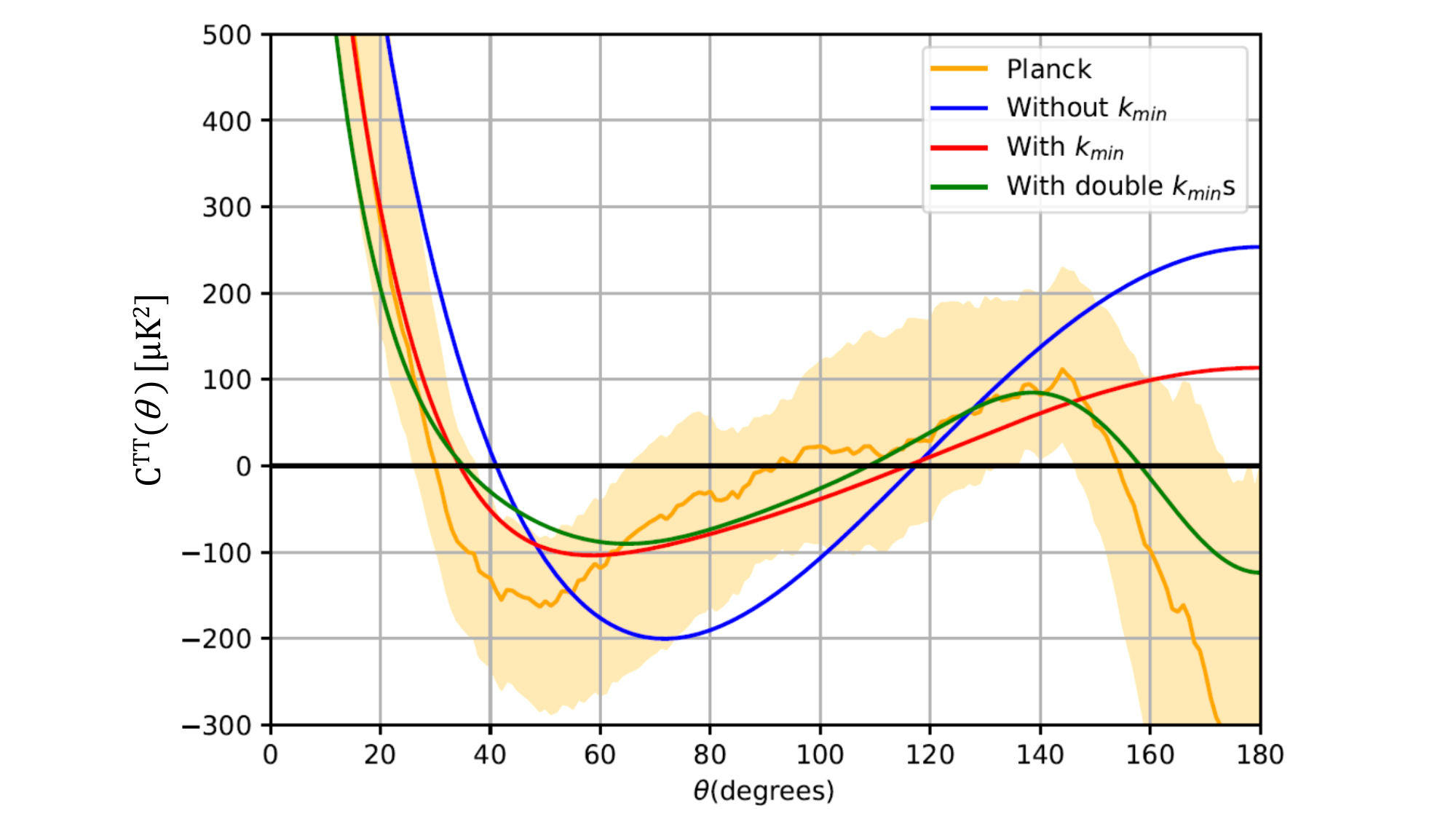}
\includegraphics[width=8.2cm]{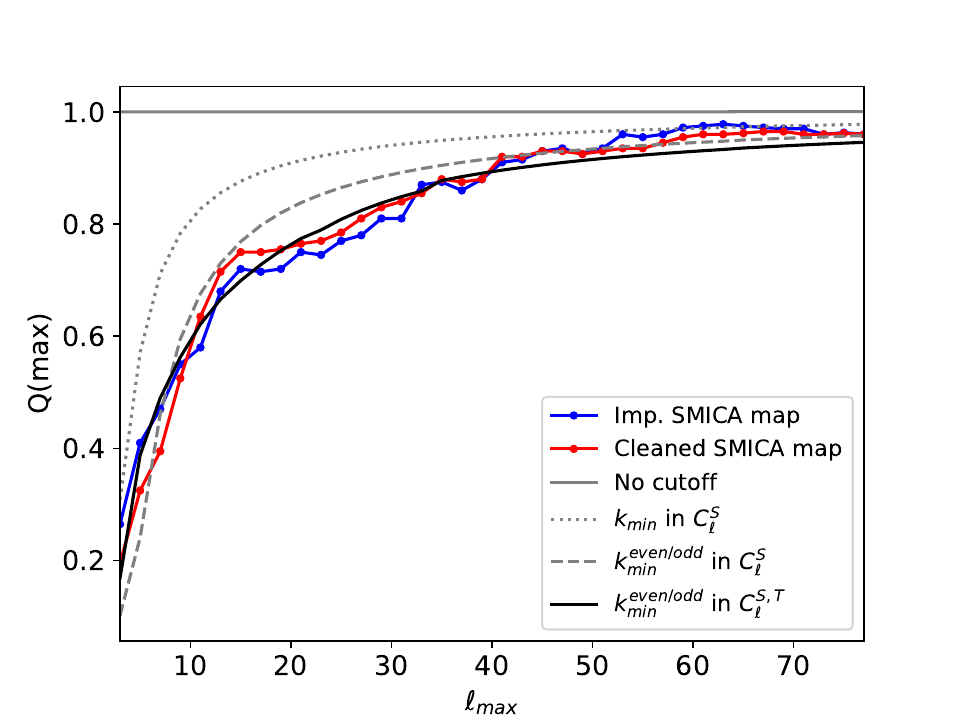}

\caption{\label{fig:1} Left panel:
Two-point angular temperature correlation function $C^{\rm TT}(\theta)$ (orange), with fits from the 2018 Planck data obtained under different assumptions about the IR cutoff(s) in the scalar power spectrum. The shaded region indicates the $1\sigma$ observational uncertainty. This plot closely follows Ref.\cite{Liu-Melia:2025}, but now also includes the curve corresponding to an IR doublet (green). 
Right panel: Parity statistic $Q(\ell_{\rm max})$ versus $\ell_{\rm max}$, computed with different SMICA masks \cite{Panda(2021)}, and fitted curves under the same assumptions as in the left panel \cite{sanchis-sanz:2024}.}

\end{figure*}

Assuming isotropy, $C^{\rm TT}(\theta)$ is usually expanded in terms of Legendre polynomials as
\begin{equation}\label{eq:C2}
C^{\rm TT}(\theta)=\ \sum_{\ell \ge 2} \frac{(2\ell+1)}{4\pi}\ C_{\ell}^{\rm TT}\ P_{\ell}(\theta)\;,
\end{equation}
where $P_{\ell}(\theta)$ is the order-$\ell$ Legendre polynomial, and the sum extends from $\ell=2$ since the monopole and dipole contributions have been removed from the analysis.

In the following, we neglect the transfer function and focus on the Sachs-Wolfe effect \cite{SW:1967} as the dominant source of primary (scalar) anisotropies, which is expected to dominate on scales larger than $\simeq 1^{\circ}$. Consequently, the multipole coefficients of Eq.~\ref{eq:C2} can be computed assuming a flat power spectrum, as
\begin{equation}\label{eq:Cellnocutoff}
C_{\ell}^{\rm TT}{(\rm scalar)}\ =\ N^S\ \int_0^{\infty}\ du\ \frac{j_{\ell}^2(u)}{u}\;,
\end{equation}
where $j_{\ell}$ is the spherical $\ell$-Bessel function, and $N^S$ stands for a normalization factor to be determined from the fit to observational data. 
The resulting curve $C^{\rm TT}(\theta)$ curve deviates significantly from the measured data (see the blue curve in Fig.~1), indicating that further action is required \cite{Melia:2012}.

\subsection{Infrared cutoff in the scalar power spectrum}

To address the above-mentioned unsatisfactory result naively expected from standard cosmology, Ref.~\cite{Melia:2018} introduced an IR cutoff $k_{\min}$ in the scalar power spectrum, which in turn implies a lower limit $u_{\min}$ in the integral Eq.\ref{eq:Cellnocutoff} of the  $C_{\ell}^{\rm TT}$ coefficient :
\begin{equation}\label{eq:Cellcutoff}
C_{\ell}^{\rm TT}{\rm (scalar)}\ =\ N^S\ \int_{u_{\rm min}}^\infty\ du\ \frac{j_{\ell}^2(u)}{u}\;,
\end{equation}
where $u_{\rm min}$ is related to the IR cutoff as
\begin{equation}\label{eq:u1}
k_{\rm min}= \frac{u_{\rm min}}{r_d}\;,
\end{equation}
 and $r_d$ denotes the co-moving distance to the last scattering surface, typically about 14,000 Mpc, setting a causality scale limit to angular correlations across the sky \cite{Hogan:2021pap}.

 In our approach, $k_{\rm min}$ denotes a sharp cutoff in the scalar power spectrum (later extended to the tensor power spectrum), which removes all Fourier modes below this scale. As discussed in the Introduction, the idea of introducing a lower cutoff in the primordial power spectrum was originally proposed in Ref.~\cite{Contaldi:2003zv}, and later applied to observational data motivated by various theoretical considerations.

 In particular, the authors of Ref.~\cite{Melia:2018} first introduced the IR cutoff in a rather heuristic manner, with the aim of removing the unobserved positive temperature correlations at large angles that are expected in standard cosmology. By fitting the $C^{\rm TT}(\theta)$ correlation function to the Planck data and truncating the power spectrum accordingly, they obtained \cite{Melia:2018,Melia:2021tvx}
\begin{equation}\label{eq:kminvalue}
u_{\rm min}=4.34 \pm 0.36\ \leftrightarrow\ 
k_{\rm min} = 3.14 \pm 0.05 \times 10^{-4}\ \rm{Mpc^{-1}}\;.
\end{equation}
A compatible result with a single cutoff was later obtained in Ref.~\cite{sanchis-lozano:2022}, incorporating into the analysis the odd-parity preference indicated by available data.

Let us remark that the above results, on their own, provide a $\gtrsim 8\sigma$ evidence for the existence of such a cutoff, which is substantially stronger than in other approaches and estimates \cite{Contaldi:2003zv,Cline:2003ve,Kuhnel:2010pp,Hazra:2014jwa,Gruppuso:2016,Ashtekar:2021izi}, where the statistical significance is much smaller.

Moreover, a study focusing on the first multipoles of the temperature angular power spectrum yielded a consistent estimate for the IR cutoff, though somewhat smaller and with larger error bars: $k_{\rm min} = 2.04^{+1.4}_{-0.79} \times 10^{-4}\ \rm{Mpc^{-1}}$.

In sum, the existence of an IR cutoff in the scalar power spectrum may be regarded as a plausible explanation for the lack of long-range angular correlations in the CMB temperature anisotropies, as well as for the odd-parity preference of the low multipoles observed (though still marginally) in the data, as mentioned in the introduction. However, as is customary in scientific research, confirmation from different and independent tests must be carried out. In this case, such tests naturally extend to polarization angular correlations, which will be addressed later.

On the other hand, $k_{\rm min}$ was physically associated with the onset of inflation in \cite{Liu:2020frj,Liu:2024mvp}. 
Replacing the single IR cutoff with a doublet, as postulated in this work, does not contradict this interpretation, since both IR cutoffs are of the same order of magnitude as the single one, all representing comparable cosmic times.

\subsection{Double infrared cutoff in the scalar power spectrum}

It should be noted that the observed odd-parity dominance \cite{Land(2005),Copi(2010),Creswell(2021),Zhao(2015)} cannot be fully reproduced with a single cutoff, since its effect on the multipole coefficients is confined mainly to the few lowest $\ell$ terms. Consequently, in Ref.~\cite{sanchis-lozano:2022s} a pair of IR cutoffs, rather than a single one, was introduced in order to amplify the differing influence of the lower integration limits on even and odd $\ell$ modes. Let us adopt the same notation and write 
\begin{equation}\label{eq:u2}
k_{\rm min}^{\rm odd/even}=
\frac{u_{\rm min}^{\rm odd/even}}{r_d} 
\end{equation}
corresponding to two IR cutoffs (instead of one) in the
scalar power spectrum, to be associated with the odd/even multipoles.

Thus we rewrite the integral of Eq.\ref{eq:u2} as
\begin{equation}\label{eq:Cellcutoffs}
C_{\ell_{\rm odd/even}}^{{\rm TT}}{\rm (scalar)}\ =\ N^S\ \int_{u_{\rm min}^{\rm odd/even}}^{\infty}\ du\ \frac{j_{\ell}^2(u)}{u}\;,
\end{equation}
where the lower limits of the integral
$u_{\rm min}^{\rm odd/even}$ (defined in Eq.\ref{eq:u2}) affect differently the odd and even multipole coefficients (only up to $\ell \lesssim u_{\rm min}^{\rm odd/even}$, respectively), thereby altering the shape of $C^{\rm TT}(\theta)$.

From a best-fit $\chi_{\rm d.o.f.}^2$ analysis of $C^{\rm TT}(\theta)$ to the {\it Planck} temperature anisotropy data, we obtained for the scalar IR cutoff \cite{sanchis-lozano:2022s}
\begin{equation}\label{eq:us}
u_{\rm min}^{\rm odd} = 2.67 \pm 0.31\ ,\ u_{\rm min}^{\rm even} = 5.34 \pm 0.62\; ,
\end{equation}
or equivalently
\begin{equation}\label{eq:ks}
k_{\rm min}^{\rm odd} = 1.93 \pm 0.22 \times 10^{-4}\ \rm{Mpc^{-1}}\ ,\ k_{\rm min}^{\rm even} = 3.86 \pm 0.44 \times 10^{-4}\ \rm{Mpc^{-1}}\; ,
\end{equation}
whose mean is not far from the result obtained in \cite{Melia:2018} shown in Eq.\ref{eq:kminvalue}.

\subsection{Incorporating tensor modes into the analysis}

In addition to the dominant scalar modes, the contribution of tensor modes to the CMB temperature anisotropies was examined in a previous paper \cite{sanchis-sanz:2024}, with particular focus on angular correlations. To improve the fit to observational data, a set of IR cutoffs was introduced in both the primordial scalar and tensor power spectra.

The corresponding multipole coefficients for temperature correlations, $C_{\ell}^{\rm TT}({\rm tensor})$ are given by \cite{Mukhanov:2005}
\begin{equation}\label{eq:Celltensoroddeven}
C_{\ell_{\rm odd/even}}^{\rm TT}{\rm (tensor)}\ =\ N^T \ 
\frac{(\ell+2)!}{(\ell-2)!}\ 
\int_{u_{\rm min}^{\rm odd/even}({\rm tensor})}^{\infty}du\ \frac{j_{\ell}^2(u)}{u^5}\;,
\end{equation}
for $\ell \ge 2 $, distinguishing odd and even modes by different
lower cutoffs, as for the scalar case.

 Despite the rather modest effect of tensor modes on temperature anisotropies, their inclusion in our analysis of angular correlations \cite{sanchis-sanz:2024} leads to a slight but noticeable improvement in the fit to the correlation data, as can be seen in both panels of Fig.~1. To examine this question in more depth, we next present a study of the parity balance among the multipoles contributing to the temperature anisotropies.

\begin{figure*}[ht]
\centering
\includegraphics[width=12.0cm]{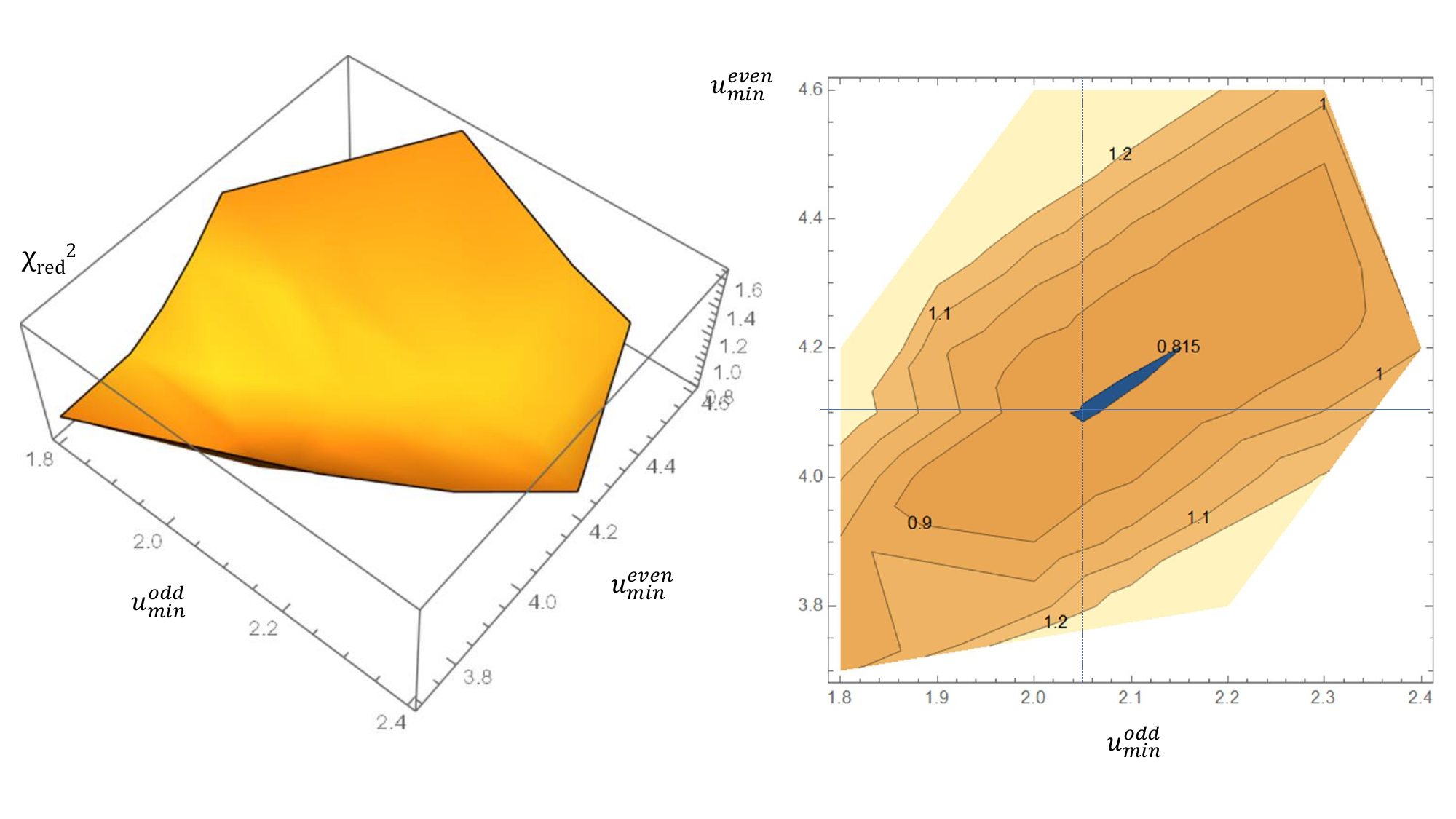} 
\includegraphics[width=12.0cm]{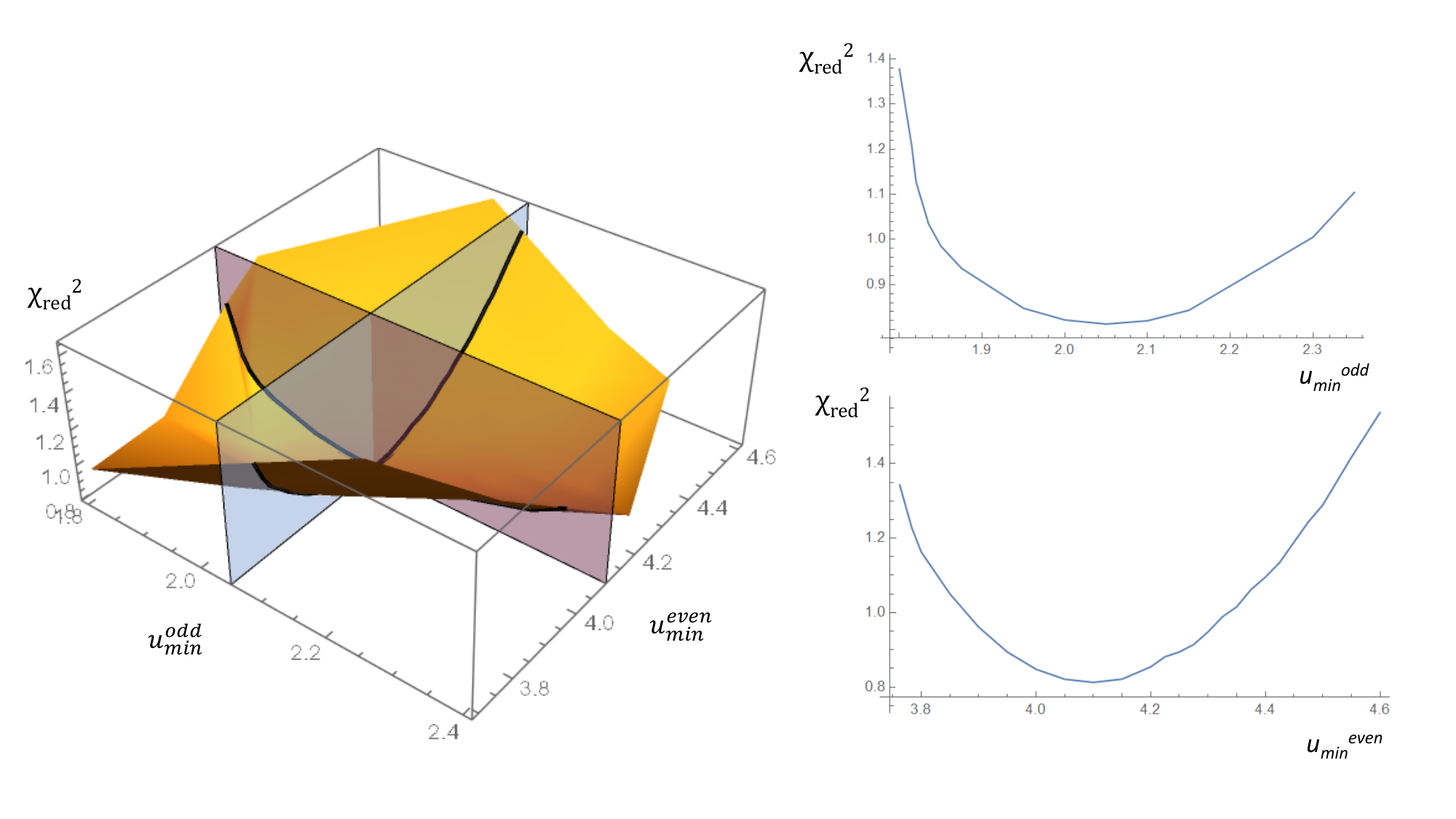}
\caption{\label{fig:2} Upper-left panel: 
3D plot of $\chi_{\rm d.o.f.}^2$ obtained from the fit of $Q(\ell_{\rm max})$
as a function of $u_{\rm min}^{\rm odd/even}({\rm scalar})$ considered now as independent variables. Upper-right panel: corresponding contour plot. Lower-left panel: planes for $u_{\rm min}^{\rm odd}=2.05$ and $u_{\rm min}^{\rm even}=4.1$, respectively; Lower-right panel: curves resulting from the their intersections with the 3D surface. The minimum was determined from the stationary point of a smooth surface fitted to the data.}
\end{figure*}

\subsection{parity-statistic study}

The deviation from even–odd parity balance in angular correlations can be quantified using a parity statistic, defined in Refs.~\cite{Aluri(2012),Panda(2021)}
\begin{equation}\label{eq:Qstat}
Q(\ell_{\rm max})=\frac{2}{\ell_{\rm max}^{\rm odd}-1}\ \sum_{\ell=3}^{\ell_{\rm max}^{\rm odd}}\ 
 \frac{D_{\ell-1}^{TT}}{D_{\ell}^{TT}}\ ;\ \ell_{max}^{\rm odd} \ge 3\;,
\end{equation}
where $\ell_{max}^{\rm odd}$ stands for the maximum odd multipole up to which the statistic is evaluated. 

Any deviation of this statistic from unity, as a function of $\ell_{\max}^{\rm odd}$, indicates an even–odd parity imbalance: values below unity imply odd-parity dominance and a downward tail at large angles in the $C^{\rm TT}(\theta)$ plot, whereas values above unity imply even-parity dominance and an upward tail. The parity statistic $Q(\ell_{\max})$ actually plays an important role in our analysis, as it helps assess the relative contributions of scalar and tensor modes to angular correlations.

The right panel of Fig.~1 (from Ref.\cite{sanchis-sanz:2024}) shows $Q(\ell_{\rm max})$ as a function of $\ell$, providing a 'zoomed-in' view of the low-$\ell$ region, or equivalently, the large-angle scale, complementing the plot in the left panel. Note the slight but noticeable improvement of the goodness of the fits 
once tensor modes are incorporated into the analysis, in addition to the scalar ones. Below, we develop further this analysis focusing on the parity imbalance, relaxing the requirement $u_{\min}^{\rm even}(\mathrm{scalar})=2u_{\min}^{\rm odd}(\rm{scalar})$ which was derived in our previous study based on the $C^{\rm TT}(\theta)$ function \cite{sanchis-lozano:2022s}.

Specifically, we present  
a scan over the $u_{\min}^{\rm odd}(\rm {scalar})$ and $u_{\min}^{\rm even}(\rm {scalar})$ plane, treating them as {\em independent variables} in the fit to $Q(\ell_{\max})$ data, while we  keep the constraint in the tensor sector: $u_{\min}^{\rm odd}(\mathrm{tensor}) = 2u_{\min}^{\rm odd}(\mathrm{scalar})$ and $u_{\min}^{\rm even}(\mathrm{tensor}) = 2u_{\min}^{\rm even}(\mathrm{scalar})$, a choice that will be justified in the following section. 

In Fig.~2 we show different 3D and 2D plots of the reduced $\chi^2$, computed for each pair of $u_{\min}^{\rm odd}(\mathrm{scalar})$ and $u_{\min}^{\rm even}(\mathrm{scalar})$ from our scan, with a step size of 0.05 for both variables. A conical-shaped surface emerges in 3D, whose minimum yields the mean values of $u_{\min}^{\rm odd/even}(\mathrm{scalar})$. This point is indicated in the lower-right panel by the intersection of the two lines superimposed on the contour plot, corresponding to a $\chi_{\rm d.o.f.}^2 \simeq 0.81$ (see contour plot of the upper-right panel).

The location of the minimum was determined using the {\it Mathematica} package by constructing a smooth representation of the ($u_{\rm min}^{\rm odd},u_{\rm min}^{\rm even}, \chi_{\rm d.o.f.}^2$) data and identifying the stationary point of this surface. The resulting coordinates were taken as the estimate of the minimum. Uncertainties were obtained by propagating the assumed measurement errors of the data (determined from different SMICA masks) through the fitting/interpolation procedure (via Monte Carlo), yielding standard errors on the minimum coordinates that are taken as error bars.

Finally, we find for the scalar infrared cutoff doublet: 
\begin{equation}\label{eq:ust}
u_{\rm min}^{\rm odd}{\rm (scalar)} = 2.05^{+0.25}_{-0.20} \ ,\  u_{\rm min}^{\rm even}({\rm scalar}) =4.10^{+0.26}_{-0.20}    
\end{equation}
which represents a small decrease compared to the values of Eq(\ref{eq:u2}) without tensor modes. This can be qualitatively understood by noting that the tensor modes contribute to the even–odd asymmetry, thereby reducing the degree of parity breaking required from the scalar sector to account for the observed odd-parity preference. 

From Eq.~\ref{eq:ust}, the ratio of the two scalar IR cutoffs is found to be exactly 2 (within the associated uncertainties), fully consistent with the result obtained in \cite{sanchis-lozano:2022} using a fit to the $C^{\rm TT}(\theta)$ correlation function.

\section{Early universe topology from a Kaluza-Klein model}

We propose a specific scenario in which the cutoffs in the angular power spectra arise naturally. Our framework is based on a simple five-dimensional (5D) space-time model that yields a four-dimensional (4D) low-energy theory when the fifth dimension is compactified to a circle of radius $R$ at an early epoch, close to the exit of the Planck era and around the Grand Unified Theory (GUT) scale \cite{sanchis-sanz:2024}. This compactification gives rise to 4D Kaluza–Klein (KK) towers whose mass spectra determine the cutoffs in the Fourier expansion of the fields relevant to our CMB analysis.

The inverse of the radius of the compactified extra dimension $1/R$ sets the lowest mass of the KK tower, acting as an IR cutoff for fields in our 4D world. On the other hand, the origin of the IR doublet could be related to the fulfillment of different combinations of (Dirichlet/Neumann) boundary conditions (BCs) \cite{sanchis-sanz:2024}, following orbifold compactification.

\subsection*{Orbifold compactification of the fifth dimension}

As previously anticipated, we postulate that the fifth extra dimension of the 5D world becomes compactified just before the onset of inflation. For the moment, we will also assume that the five-dimensional spacetime is described by a flat geometry, specifically with a 5D Minkowski metric:
\begin{equation}\label{eq:metric}
    ds^2 = dt^2-dx_i dx^i - dy^2 \;,
\end{equation}
where $i$=1, 2, 3, while $y$ denotes the 5-th dimension. Following orbifold compactification \cite{Csaki:2004ay}, the fifth dimension ranges over the interval $y \in [0,L]$, where $L=\pi R$ (see Fig.3)· 
Only scalar fields, and no fermionic fields \cite{vonGersdorff:2004eq}, will be considered in our approach.

Scalar fields  propagating in this geometry can be factorized in the so-called KK decomposition as
\begin{equation}\label{eq:5D}
\Phi(x^\mu, y) = \sum_n \phi_n(x^\mu) \, f_n(y)\;,
\end{equation}
where the 4D field $\phi(x^\mu)$ satisfies the Klein-Gordon equation of motion, while $f(y)$ satisfies a wavefunction equation which depends on the geometry and BCs at both ends of the interval.
Under certain assumptions, the parity of the KK profile $f_n(y)$ determines the 3D parity of the corresponding 4D component $\phi(x^\mu)$, making it a key element in our analysis.
Our later development can proceed in two ways (both leading to the same physics):
\begin{itemize}
\item orbifold $ S^1/(Z_2 \times Z'_2)$ compactification requiring invariance under some parity operations in the fifth dimension. 
\item Neumann/Dirichlet BCs at both ends of the compactified dimension. 
\end{itemize}

\begin{figure*}[ht]
\centering
\includegraphics[width=11.0cm]{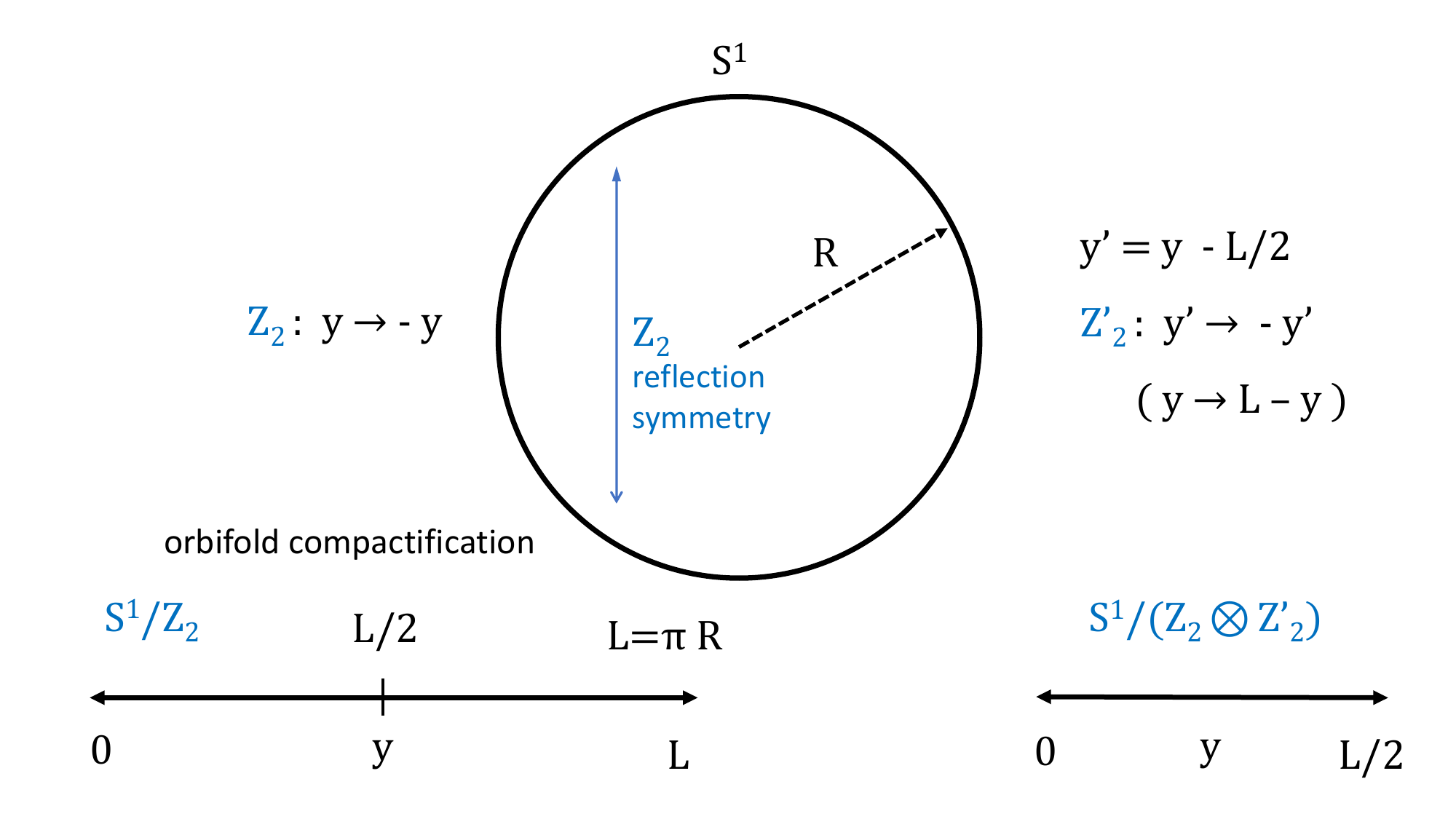}
\caption{\label{fig:3} Sketch of orbifold compactification of a circular extra dimension of radius $R$ (basically determined by the GUT energy scale in our case), showing the effect of $S^1/(Z_2 \times Z'_2)$ symmetry. See the main text for a more detailed explanation of the different symmetry transformations and associated parities.}
\end{figure*}

Indeed, by defining additional parities on the compactified extra dimension
beyond the symmetry: $y+2\pi R$, distinct topologies are obtained, ultimately leading to observational consequences in the 4D world. Here, we will focus on a minimal $S^1/(Z_2 \times Z'_2)$
orbifold in a 5D space-time, which has been already applied to the GUT epoch \cite{Hebecker:2001j}:  

\begin{itemize}

\item  $Z_2$: reflection $y \to -y$, with a fixed point at $y=0$

\item  $Z'_2$: defining $y'=y-\pi R/2$, followed by $y' \to -y'$ amounts to the reflection $y \to \pi R-y$, with a fixed point at $\pi R/2$.

\end{itemize}

In our setup, the combined $ Z_2 \times Z'_2$ symmetry provides a global discrete label for the 4D fields, ultimately determining the 3D spatial parity of the 4D scalar and tensor modes, and, in turn, their correspondence to even and odd multipoles in the expansion of the angular correlation functions.

\section{Boundary conditions on a flat geometry}

Alternatively, one can define a theory directly on the segment $y \in [0,L]$, where $L=\pi R$. By choosing appropriate Neumann 
 and Dirichlet BCs at the endpoints, one can mimic the mode spectrum
 and parity structure of the orbifold compactification yielding the same physics. Let us write down the KK mode wavefunctions $f_n(y)$ for a scalar field compactified on a segment $y \in [0,L]$, with $L=\pi R$ under the four possible combinations of Neumann (N) and Dirichlet (D) BCs at the end points.

 \begin{itemize}    
 
\item Neumann-Neumann (+,+) with BCs: $\partial_yf_n(0)=0,\ \partial_yf_n(L)=0$.

The set of solutions and allowed mass spectrum (including zero mode) can be written as:
\begin{equation}\label{eq:NN}
f_n(y)=\cos{ \biggl( \frac{2n y}{R}\biggr)}=\cos{ \biggl( \frac{2n \pi y}{L} \biggr)},\ m_n=\frac{2n}{R},\ n=0,1,2,\cdots
\end{equation}

Note that all modes are even under both 
$y \to -y$ and $y \to L-y$. The combined parity turns out to be {\em even}.

\item Dirichlet-Dirichlet ($-$,$-$) with BCs: $f_n(0)=0$, $f_n(L)=0$

Solutions and mass spectrum (no zero mode):
\begin{equation}\label{eq:DD}
f_n(y)=\sin{ \biggl( \frac{(2n+2) y}{R} \biggr)}=\sin{ \biggl( \frac{(2n+2) \pi y}{L} \biggr)},\ \ m_n=\frac{(2n+2)}{R},\ n=0,1,2,\ldots
\end{equation}
The combined parity is {\em even}.

\item Neumann-Dirichlet (+,$-$) with BCs: $\partial_yf_n(0)=0$,  $f_n(L)=0$

Solutions and mass spectrum (no zero mode):
\begin{equation}\label{eq:ND}
f_n(y) =  \cos{ \biggl(\frac{(2n+1) y}{R} \biggr)} =  \cos{ \biggl(\frac{(2n+1) \pi y}{L} \biggr)},\ m_n=\frac{(2n+1)}{R},\ n=0,1,2,\ldots
\end{equation}
The combined parity is {\em odd}.

\item Dirichlet-Neumann ($-$,+) with BCs: $f_n(0)=0$, $\partial_yf_n(L)=0$

Solutions and mass spectrum (no zero mode):
\begin{equation}\label{eq:DN}
f_n(y) = \sin{ \biggl(\frac{(2n+1) y}{R} \biggr)} =\sin{ \biggl(\frac{(2n+1) \pi y}{L} \biggr)},\ m_n=\frac{(2n+1)}{R},\ n=0,1,2,\ldots
\end{equation}
The combined parity is {\em odd}.

\end{itemize}

Thereby, the mass spectrum of scalar fields with BCs ($-$,$-$) and ($-$,+) can be related from Eqs.\ref{eq:DD} and \ref{eq:DN} as follows,
\begin{equation}
\frac{m^{(-,-)}_n}{m^{(-,+)}_n}=\frac{2n+2}{2n+1}\;,\ n=0,1,2,\ldots\;.
\end{equation}

For 5D fields, the lowest mass mode in their KK spectrum ($n=0$) provides a natural infrared (IR) cutoff for their physical behavior in 4D. Accordingly, the IR cutoffs represented by $k_{\rm min}^{\rm odd/even} (\mathrm{scalar})$ are simply related by a factor 2,
\begin{equation}\label{eq:ratiokscalar}
    k_{\rm min}^{\rm even} {\rm (scalar)}=2\ k_{\rm min}^{\rm odd} {\rm (scalar)}\;.
\end{equation}

This ratio coincides with the value obtained for scalar modes in our phenomenological analysis of temperature angular correlations in the CMB. In a previous study \cite{sanchis-lozano:2022s}, such a “magical” value of 2 was interpreted in terms of periodic and antiperiodic boundary conditions satisfied by a Dirac field.

\subsection*{Tensor modes}

As previously noted, tensor modes of the CMB—generated by metric perturbations in the early universe, but imprinted much later, contribute to both temperature and polarization anisotropies, albeit to different extents. Despite their distinct origins and characteristics, scalar and tensor perturbations will be treated within a unified theory.

Indeed, spin-2 fields are unavoidably present in any KK spacetime geometry. From the four-dimensional perspective, the KK decomposition of the spin-2 graviton field includes a massless mode, responsible for mediating 4D gravity, together with a tower of massive excited states (the KK gravitons).  As before, the lowest massive state sets the IR cutoff in the 4D world..

By imposing the same boundary conditions as for the scalar modes, one can write down the following ratio of the lowest massive modes with different combined parity, according to Eqs.\ref{eq:NN} and \ref{eq:ND}, as
\begin{equation}\label{eq:eq:kratioscalar}
\frac{m_1^{(+,+)}}{m^{(+,-)}_0}=2\;,
\end{equation}
where $m_1^{(+,+)}$ denotes the first excited (non-zero mass, $n=1$) KK state.

Thus, we can write for the tensor IR cutoffs a relation equivalent to
Eq.\ref{eq:ratiokscalar}, 
\begin{equation}
 k_{\rm min}^{\rm even} {\rm (tensor)}=2\ k_{\rm min}^{\rm odd} {\rm (tensor)}\;,
\end{equation}
where their numerical values are again basically set by $1/R$. Therefore
one naively expects: $k_{\rm min}^{\rm odd/even}{(\rm scalar)} \simeq k_{\rm min}^{\rm odd/even}{(\rm tensor)}$ in flat geometries. However, still there is a way
of modifying the masses of tensor modes (and hence the IR cutoffs and their ratios) in warped geometries, as discussed below.

At this point, a caveat is in order: we have employed a simple KK model throughout, with specific assignments of IR cutoff ratios based on certain parity properties, intended for later use in interpreting physical results on angular correlations. The aim of this Section has been to demonstrate that such assignments are indeed feasible, but not to develop a detailed model of the early universe.

\section{Warped geometries}

The above construction can be generalized to metrics with curvature in the extra-dimension, commonly called {\it warped} geometries \cite{Randall:1999ee,Randall:1999vf,Csaki:2004ay}, employed in different fields and applications (see, e.g., \cite{Maldacena:1997re,Hirn:2005nr,Donini:2021kne,Hirn:2006nt,Hirn:2006wg}.

Warped geometries can be parametrized in terms of a warp factor $w(z)$ in a  factorizable metric
\begin{equation}
     ds^2 = w(z)^2 (dt^2-dx_i dx^i - dz^2)\;,
\end{equation}
where conformal coordinates have been used. For example, $w(z)= 1/z$ in the Randall-Sundrum model \cite{Randall:1999ee}.

The following sum rule allows one to obtain the behavior of the even and odd-parity KK towers in warped extra-dimensions for scalar modes,
\begin{equation}\label{eq:sumrules}
    \frac{k_{\rm min}^{\rm even}}{k_{\rm min}^{\rm odd}} \simeq \frac{\int_{z_0}^{z_1} w(z) \alpha(z) dz \int_{z}^{z_1} d z'/w(z')}{ \int_{z_0}^{z_1} w(z) \alpha(z) dz \int_{z_0}^{z} d z'/w(z') }\;,
\end{equation}
where $z_0$ and $z_1$ are related to the AdS curvature and the warp factor; 
$\alpha(z)=\frac{\int_{z}^{z_1} d z'/w(z')}{\int_{z_0}^{z_1} d z'/w(z')}$, see~\cite{Hirn:2007bb} for more details. Using this sum rule, we recover the relation of Eq.~\ref{eq:sumrules} for the flat metric case, where $w(z)=1$. In general, the ratio 
$$k_{\rm min}^{\rm even}/ k_{\rm min}^{\rm odd} \gtrsim 2$$ 
continues to hold for non-flat geometries, so we write for scalar modes
$k_{\rm min}^{\rm even}/k_{\rm min}^{\rm odd}\ =\ 2q$, or equivalently
$$u_{\rm min}^{\rm even}({\rm scalar})/u_{\rm min}^{\rm odd}({\rm scalar})\ =\ 2q$$
for computation of the multipole coefficients. Note that $q$ captures the influence of the warped geometry on the mass spectrum (in particular on the lowest excited states), such that $q=1$ for a flat geometry while $q\gtrsim 1$ is typically expected for the warped ones. A relative small $q$ value is expected in mild warped extra-dimensions
\cite{McDonald:2008,Medina:2010mu}.

\subsection*{Tensor modes}

Tensor modes would display a similar behavior in 
the presence of curvature, as explained in Refs.~\cite{Randy,Dillon}, so that we shall write 
$k_{\rm min}^{\rm even}/ k_{\rm min}^{\rm odd}\ =\ 2\tilde{q}$, or equivalently
$$u_{\rm min}^{\rm even}({\rm tensor})/u_{\rm min}^{\rm odd}({\rm tensor})\ =\ 2\tilde{q}$$

Moreover, another hierarchy between the tensor and scalar IR cutoffs can be established: 

$Q=k_{\rm min}^{\rm even}({\rm tensor})/k_{\rm min}^{\rm even}({\rm scalar})$, 
where $Q=1$ corresponds to a flat geometry. Similarly, one can define
$\tilde{Q}=k_{\rm min}^{\rm odd}({\rm tensor})/{k_{\rm min}^{\rm odd}({\rm scalar})}$
Depending on the model, 
$Q$ or/and $\tilde{Q}$ could be smaller or larger than unity. 

To summarize, the following relations can be established:
\begin{eqnarray}\label{eq:q}
k_{\rm min}^{\rm even}({\rm scalar}) & = & 2q\ k_{\rm min}^{\rm odd}({\rm scalar})\ , 
\\
k_{\rm min}^{\rm even}({\rm tensor}) & = & 2\tilde{q}\ k_{\rm min}^{\rm odd}({\rm tensor})\ ,
\\
k_{\rm min}^{\rm even}({\rm tensor}) & =  & Q\ k_{\rm min}^{\rm even}({\rm scalar}) \\
k_{\rm min}^{\rm odd}({\rm tensor}) & =  & \tilde{Q}\ k_{\rm min}^{\rm odd}({\rm scalar})\;,
\end{eqnarray}
where, as already commented, $q=\tilde{q}=1$ applies to flat geometries, while $q \gtrsim 1$ and/or  $\tilde{q} \gtrsim 1$ applies to warped geometries. It is interesting to highlight the relation between the above ratios:
\begin{equation}\label{eq:R}
\tilde{q}=\frac{Q}{\tilde{Q}} \times q
\end{equation}

Any observational signature of the postulated topology
via angular correlations would require $\tilde{q} \gtrsim 1$. This condition is easily achieved if the behaviors of both $Q$ and $\tilde{Q}$ remain similar when transitioning from a flat geometry to a warped one.

\section{Even and odd multipole contributions to the two-point angular correlation function}

In light of the discussion in Sections 4 and 5, let us reexamine the association between the odd/even multipoles (arising in the expansion of the two-point correlation function in Eq.~\ref{eq:C2}) and the corresponding odd/even IR cutoffs, as well as their subsequent extension to B-mode polarization.

To this end, as previously mentioned, we assume for simplicity a 5D scalar field $\Phi(x^{\mu},y)$, whose 4D component $\phi(x^{\mu})$ represents a scalar inflaton field. Accordingly, the parity of a 4D KK mode under 3D spatial reflections $\vec{x} \to -\vec{x}$ must satisfy:
\begin{equation}\label{eq:PKK}
P_{4D}\ =\ P_{5D}\ \cdot P_{KK}\;,
 \end{equation}
where $P_{5D}$ is the intrinsic spatial parity of the 5D field,
$P_{KK}$ is the  orbifold parity of the KK wavefunction $f(y)$, and $P_{4D}$ is the resulting 3D parity of the 4D field mode $\phi_n(x^{\mu})$. Thus, the relationship between the parity of the multipole terms and the corresponding IR cutoffs used in computing the coefficients $C_{\ell_{\text{even/odd}}}$ is well justified.

\begin{figure*}[ht]
\centering
\includegraphics[width=11.5cm]{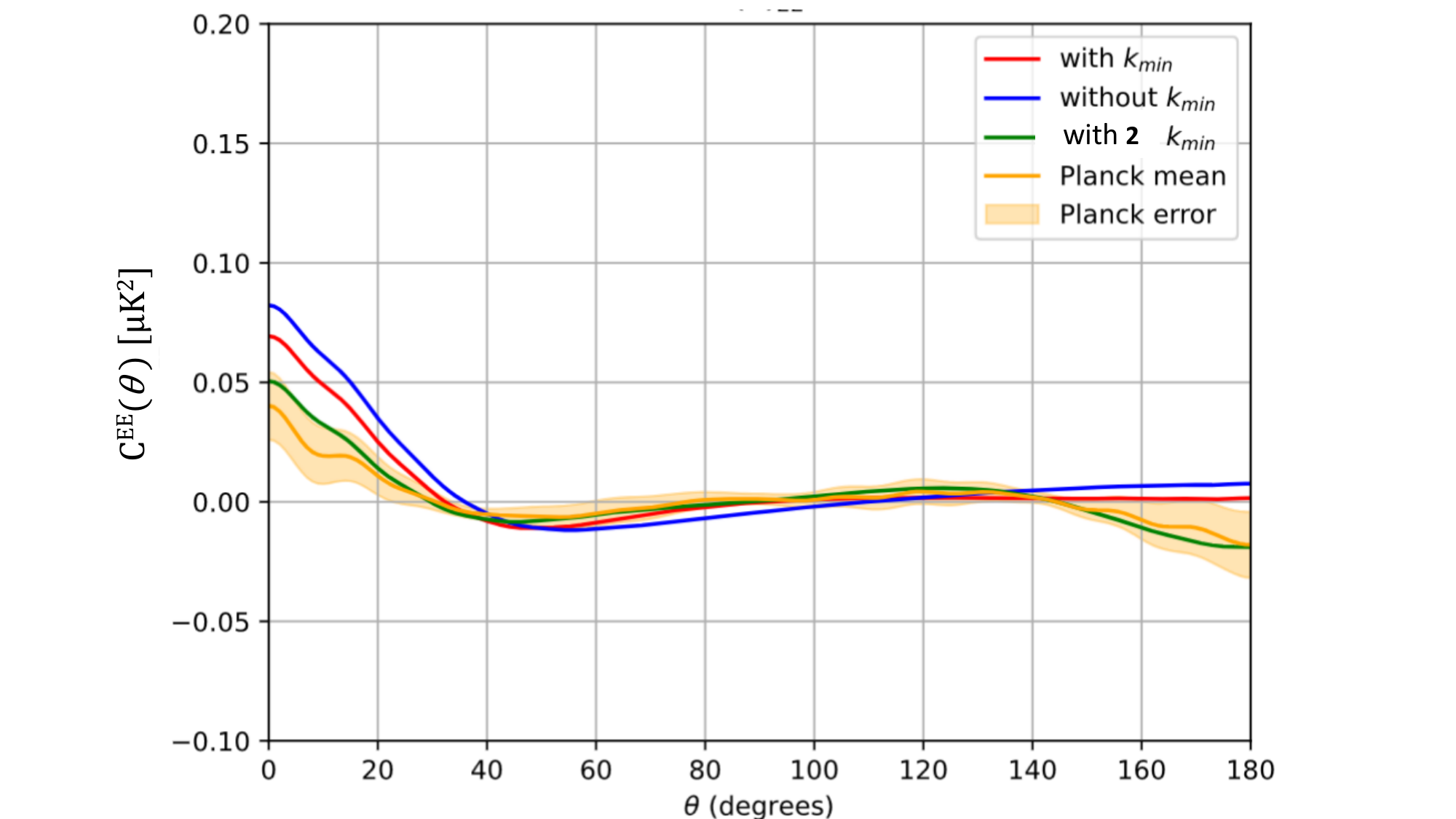}
\caption{\label{fig:4}  E-mode polarization two-point correlation function under different assumptions about the IR cutoff(s), and compared to the observed curve from {\it Planck} for $\ell < 30$ with its 1$\sigma$ uncertainty band (orange). Notice the downward tail in the case of the IR doublet (green curve), which is absent in the no-cutoff and single-cutoff cases, similar to what is observed in temperature correlations.}
\end{figure*}

\section{Polarization of the CMB versus Early-Universe Topology}

So far, our analysis has focused on the temperature anisotropies of the CMB, including the contribution of tensor modes to the angular correlation function. In what follows (the main focus of this work) we turn to the polarization signal, examining both the E-mode and B-mode components, with particular emphasis on the low-$\ell$ multipoles.

As is well known, only PGWs can generate primordial B-mode polarization in the CMB that survives to the present day, whereas other sources of non-cosmological origin contribute to a significant contamination background. Ideally, once the effects of gravitational lensing are removed from the data, the detection of B-modes would serve as a smoking gun for the existence of PGWs, and possibly for inflation itself. In contrast, scalar (density) perturbations produce only E-mode polarization at linear order in perturbation theory and therefore do not leave a direct imprint of PGWs on the CMB. Nonetheless, we emphasize that the analysis of CMB E-mode polarization, particularly at low multipoles ($\ell$), can also yield valuable insights into the very early universe, as we discuss below.

\subsection{E-mode polarization}

E-mode polarization in the CMB arises primarily from Thomson scattering of photons off free electrons in the presence of quadrupole anisotropies. The dominant contribution comes from density perturbations at recombination, while reionization acts as a second round of Thomson scattering that generates additional E-mode polarization, especially on large angular scales. Gravitational lensing also contributes to the overall E-mode signal.

Similarly to the study of temperature anisotropies, the angular correlation function for the E-mode polarization may be written as
\begin{equation}\label{eq:CEE}
C^{\rm EE}(\theta)=\ \sum_{\ell \ge 2} \frac{(2\ell+1)} {4\pi}\frac{(\ell+2)!}{(\ell-2)!}\ C_{\ell}^{\rm EE}\ P_{\ell}(\theta)
\end{equation}
The $(\ell+2)!)/(\ell-2)!$ factor has been explicitly extracted from the coefficient definition to clearly highlight that polarization correlations are highly dominated by large-$\ell$ multipoles, thereby diminishing the impact of any IR cutoff. Nonetheless, still useful information corresponding to
large angular scales, i.e. low $\ell$, can be obtained from Eq.\ref{eq:CEE}.

For this purpose, we show in Fig.~4 the $C^{\rm EE}(\theta)$ curve and its $1\sigma$ error band, derived from {\it Planck} data using the CAMB code \cite{Lewis:2000} for $\ell_{\rm max} = 30$, keeping the same format as in Ref.\cite{Liu-Melia:2025}. 
Different theoretical curves are plotted under different assumptions regarding the IR cutoffs previously obtained from the fit to temperature correlations. The difference lies now in that there is an extra curve for the IR doublet, yielding a reduced $\chi^2$ close to unity in the latter case. Qualitatively, the downward tail in the IR doublet (absent in the no-cutoff and single-cutoff cases) resembles what is seen in the temperature correlations.

\begin{figure*}[ht]

\centering
\includegraphics[width=8.0cm]{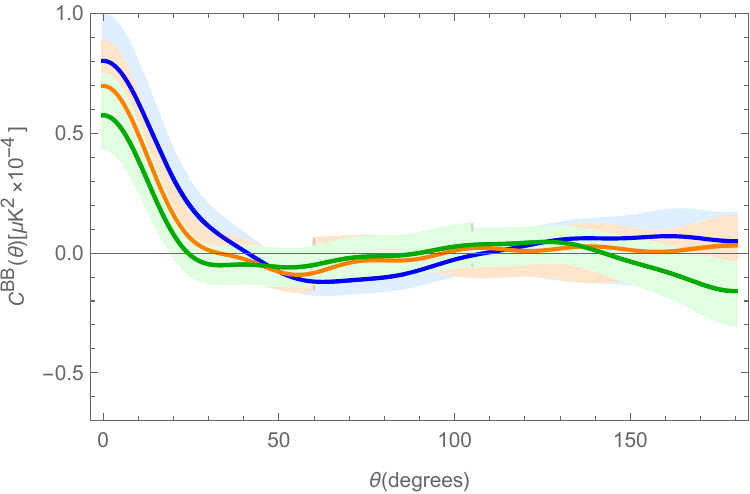}\hspace{0.3cm}
\includegraphics[width=8.0cm]{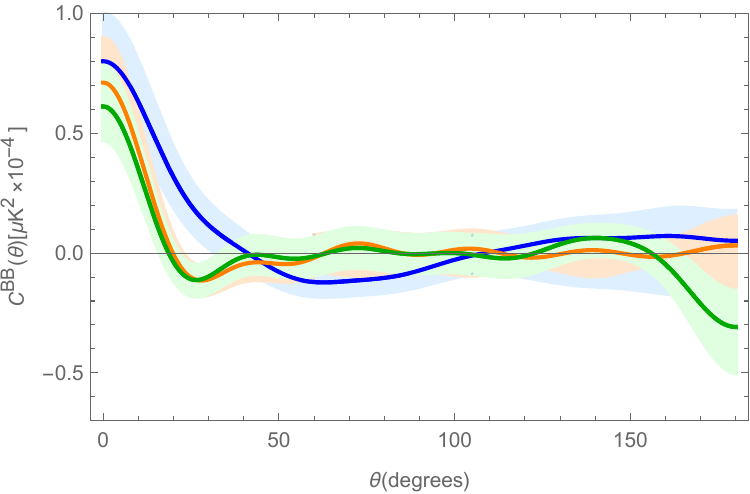}
\includegraphics[width=8.0cm]{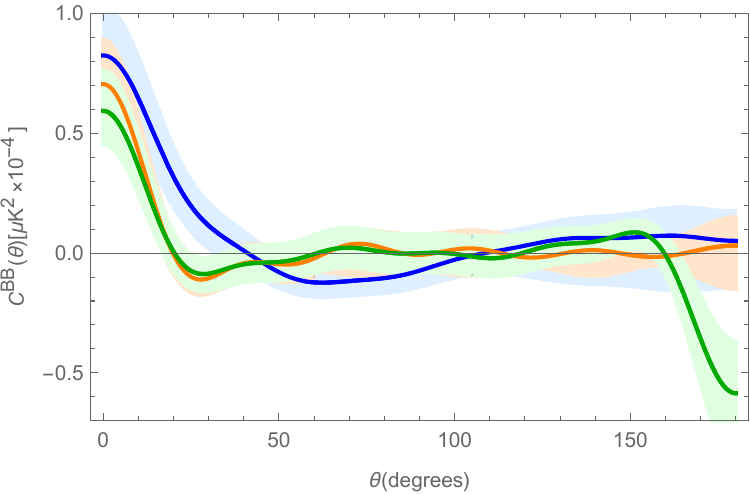}
\hspace{0.3cm}
\includegraphics[width=8.0cm]{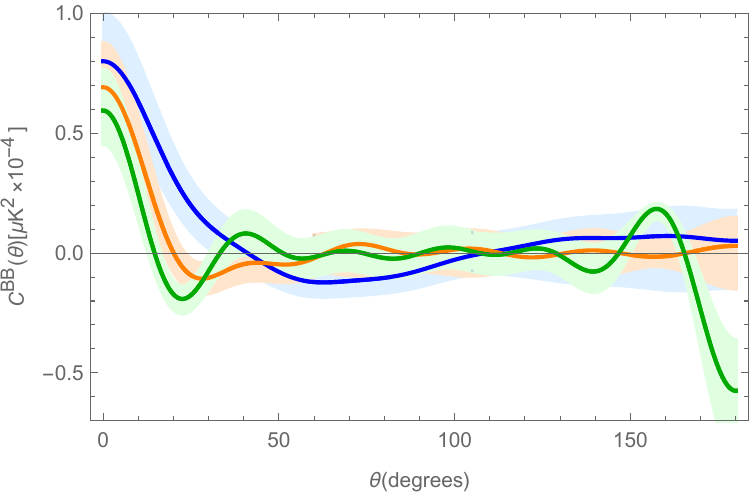}

\caption{\label{fig:5} B-mode polarization two-point correlation function $C^{\rm BB}(\theta)$ (in $10^{-4} \times \mu K^2$) 
using different assumptions on the IR cutoff(s) for $\ell \leq 11$. Blue: no cutoff, ted: one cutoff, green: two cutoffs.
Upper left panel: $u_{\rm min}^{\rm odd}=2$, $u_{\rm min}^{\rm even}=4$; Upper right panel: $u_{\rm min}^{\rm odd}=4$, $u_{\rm min}^{\rm even}=8$.
Lower left panel: $u_{\rm min}^{\rm odd} =4$, $u_{\rm min}^{\rm even}=16$; Lower right panel: $u_{\rm min}^{\rm odd}=8$, $u_{\rm min}^{\rm even}=16$. The error bands correspond to cosmic variance, assuming an ideal {\it LiteBIRD} performance without instrumental noise or gravitational lensing. Note that large values of $u_{\rm min}^{\rm even}$ suppress the even multipoles, thereby leading to odd-parity dominance.}
\end{figure*}

A more detailed study of the EE power spectrum, in addition to the two-point correlation function, would be crucial to confirm the improvement obtained by incorporating the IR doublet as a new contribution. A thorough investigation of this aspect is left for future work. On the other hand, since scalar modes likely dominate over tensor modes in the E-mode polarization, the curves in Fig.~4 are largely insensitive to $r$, which motivates the analysis of B-mode polarization in the next section.

\subsection{B-mode polarization}

B-mode polarization is often regarded as the smoking gun for PGWs and, by extension, for inflation itself. Upcoming high-precision measurements, such as those expected from the {\it LiteBIRD} \cite{LiteBIRD:2022cnt}, are anticipated to provide invaluable insights into the early universe, helping to discriminate among different cosmological models of the inflationary epoch. 

Not all sources of CMB B-mode polarization arise from PGWs. Gravitational lensing of E-modes, generated by density perturbations, converts part of their power into B-modes, with a spectrum peaking at large multipoles. In addition, polarized foregrounds such as thermal dust emission and synchrotron radiation from relativistic electrons in the Galactic magnetic field also contribute. These foreground signals, however, exhibit a frequency dependence that differs significantly from that of the primordial signal, likely enabling their mitigation through multi-frequency observations \cite{LiteBIRD:2022cnt}. 

In Refs.~\cite{Liu:2024mvp,Liu-Melia:2025}, the authors studied in detail the impact of a truncated primordial power spectrum on the CMB B-mode polarization, considering only the case of a single IR cutoff in comparison with the no-cutoff scenario. As mentioned earlier, we now extend this analysis by investigating the influence of multiple IR cutoffs on the B-mode angular correlations, focusing on large angular scales where the direct tensor contribution is most significant.

\begin{figure*}[ht]
\centering
\includegraphics[width=8.2cm]{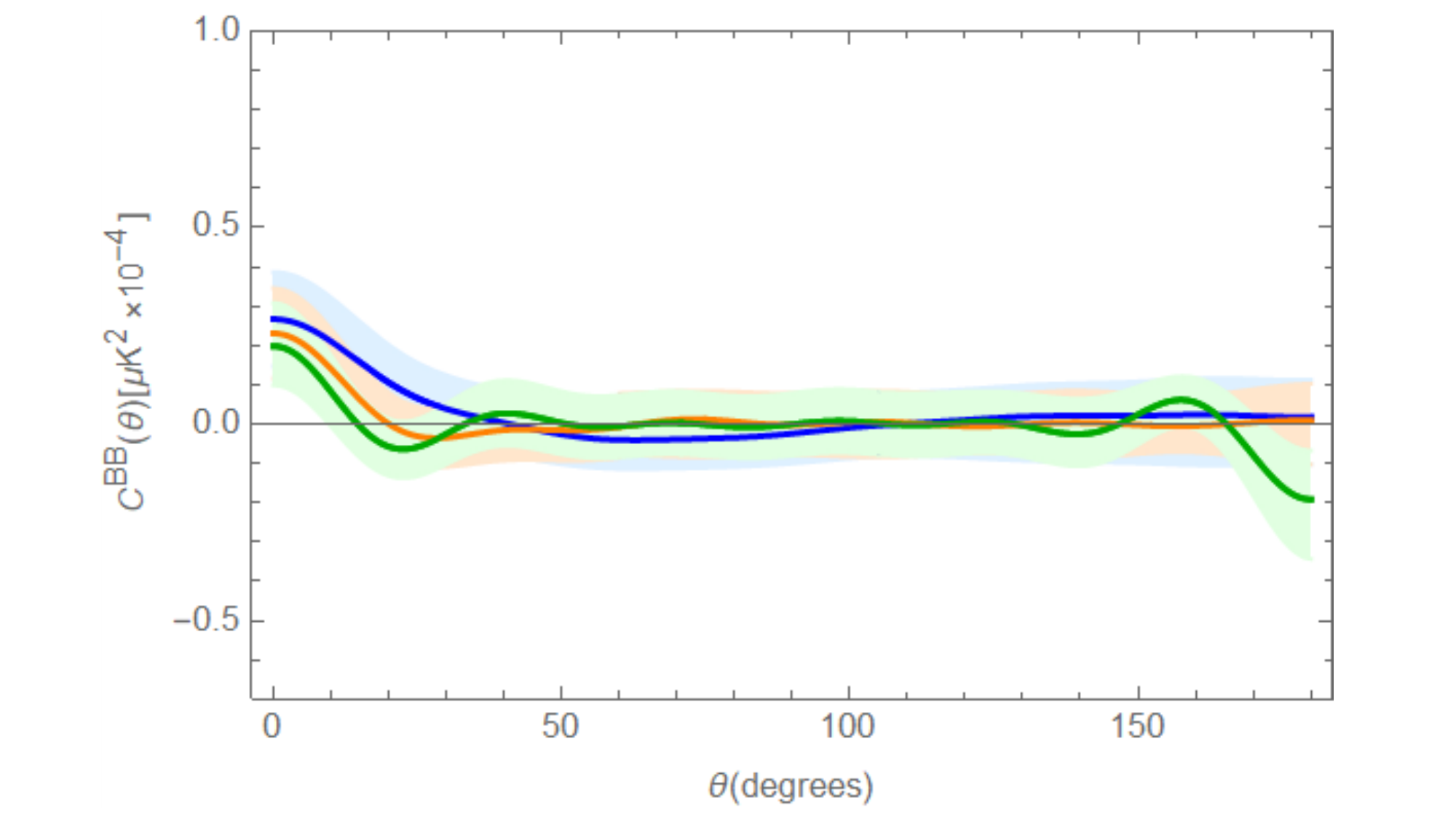}
\includegraphics[width=8.2cm]{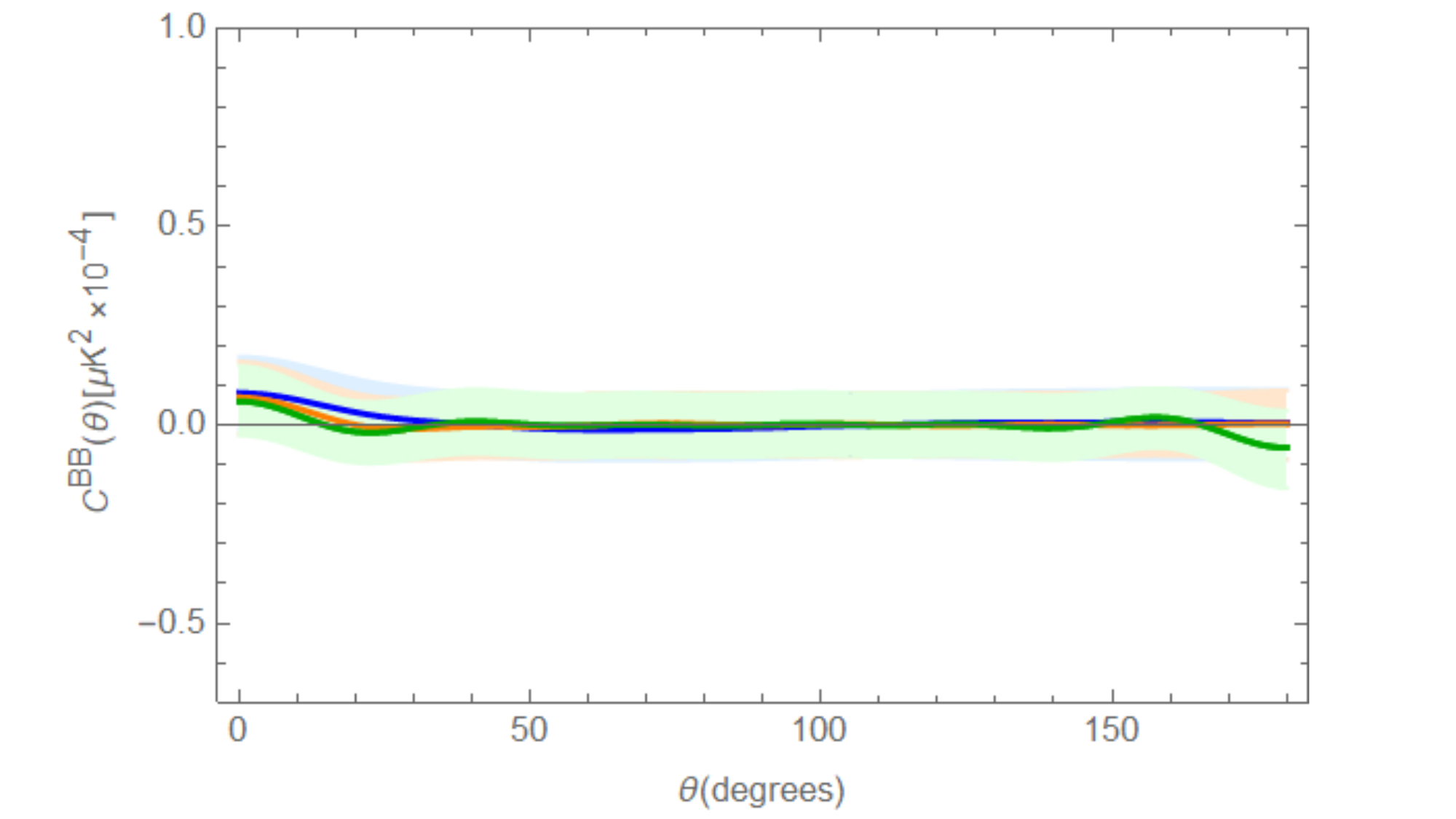}
\caption{\label{fig:6} The same as in Fig.~5 but for smaller values of $r$. Left-panel: $r=0.001$; Right-panel: $r=0.0003$. For $r < 0.001$, the three uncertainty bands overlap, which removes any ability to discriminate between them.}
\end{figure*}

As already done for E-mode polarization, we expand the BB two-point correlation function in terms of Legendre polynomials as
\begin{equation}\label{eq:CBB}
C^{\rm BB}(\theta) = \sum_{\ell \ge 2} \frac{(2\ell+1)}{4\pi} \frac{(\ell+2)!}{(\ell-2)!}\  C_{\ell}^{\rm BB}\  P_{\ell}(\theta)\;,
\end{equation}
where again the $\frac{(\ell+2)!}{(\ell-2)!}$ factor has been extracted. To avoid high multipoles, our analysis focuses on small $\ell$ values, specifically $2 \leq \ell \leq 11$ ensuring an equal number of odd and even multipoles. Moreover, beyond this range, contributions from non-cosmological sources, such as gravitational lensing, start to become dominant and are therefore removed.

In Fig.~5, we present a set of plots of $C^{\rm BB}(\theta)$ (for $2 \leq \ell \leq 11$) as a function of $\theta$, for various representative values of $u_{\rm min}^{\rm even/odd}$ (tensor) and a fixed tensor-to-scalar ratio of $r = 0.003$. In this small $r$ range, the curves shown in this figure are essentially proportional to $r$ so that they can be easily scaled for different values of $r$.

Each plot includes curves corresponding to the three scenarios considered in this study involving the primordial power spectra: (a) no IR cutoff (blue), (b) a single IR cutoff (red), and (c) two IR cutoffs (green); 1-$\sigma$ error bands have been estimated assuming ideal {\it LiteBIRD} performance, such that the uncertainties arise solely from cosmic variance. Needless to say, other statistical and systematic sources
like instrumental noise, $1/f$ noise, foreground contamination, etc,  would broaden the uncertainty bands.

In the absence of real data, one can only speculate that fitting the entire curve $C^{\rm BB}(\theta)$ together with the well-known $S_{1/2}$ statistical estimator \cite{Spergel:2003}, could reveal a distinctive signature, given the notable departure from standard expectations especially at large-angle scales.  We now take a step forward by suggesting a possible discriminating procedure.

Upon inspection of the $C^{\rm BB}(\theta)$ curves, distinct features emerge, particularly at large angular scales. A large value of $u_{\rm min}^{\rm even}$ effectively suppresses the contribution from even multipoles, causing the behavior of $C^{\rm BB}(\theta)$ to be dominated by odd multipole contributions, analogous to the effect observed in the temperature correlations shown in Fig.1. Furthermore, a large value of $\tilde{q}$ ($\tilde{q} \gg 1$), defined in Section 5, leads to a pronounced downward tail that could serve as an indication of a warped geometry. In addition, the emergence of an almost symmetric profile about $90^{\circ}$ may provide another potential signature.

The eventual detection of such a pattern in B-mode polarization data would lend strong support to the proposal put forward in this and previous works \cite{sanchis-lozano:2022s,sanchis-sanz:2024}, which aims to probe the topology of the early universe through CMB angular correlations.

Let us remark that the angular correlation function essentially contains the same information as the power spectrum, since both are derived from the same set of multipole coefficients. The correlation function, however, can provide a clearer perspective in the large-angle (low-$\ell$) regime, where the odd-parity preference emerges as a cumulative effect of several odd and even multipoles, providing a distinctive discriminating pattern.

In particular, the value of the BB correlation function in the antipodal region, i.e. $C^{\rm BB}(\theta \approx 180^{\circ})$, could be crucial
to discriminate among different topologies, as it reflects the parity preference of all contributing multipoles to the (green) curve shown in Fig.~5. 

This approach also permits the inference of a lower sensitivity limit depending on the values of $r$ and taking into account the expected uncertainty band at large angle. To this end, we present in Fig.~6 a couple of plots of $C^{\rm BB}(\theta)$ using $u_{\rm min}^{\rm odd}({\rm tensor})=8$ and $u_{\rm min}^{\rm even}({\rm tensor})=16$, for $r=0.001$ (left panel), and
$r=0.0003$ (right panel), respectively. As already commented, the error band has been computed as due purely to cosmic variance and therefore is proportional to $r$. As shown in Fig. 6, the three uncertainty bands overlap for $r < 0.001$, which eliminates any ability to discriminate between them. Moreover, uncertainties from other sources will contribute further to the error band, as already mentioned, thereby increasing its width.

Thus, we conservatively conclude that for values below $r \simeq 10^{-3}$ the pattern displayed in the plots of Fig.~5 for an IR doublet in the tensor power spectrum would be unlikely to be observed.

\section{ Discussion and conclusions}

In this paper, we first reviewed the noticeable deviation of the observed temperature two-point angular correlation function, $C^{\rm TT}(\theta)$, from the predictions of the standard cosmological model, with particular emphasis on large angular scales where the slope turns markedly toward negative values, as shown in Fig.~1.
 
Building on previous work \cite{sanchis-lozano:2022,sanchis-sanz:2024}, we further developed a possible physical origin of this phenomenon, based on a specific topology of the early universe, now extended to polarization correlations. In particular, we employed a model featuring a Kaluza–Klein setup, compactified during the GUT epoch and subject to Neumann and Dirichlet boundary conditions on an extra dimension. As a result, doublets of scalar and tensor IR cutoffs naturally emerge from the KK tower(s), associated with the even- and odd-parity transformations of multipoles in the Legendre expansion of the two-point correlation functions.

Admittedly, a simple KK model was invoked, while many other (and more elaborated) possibilities remain open, either in flat and warped geometries. Nevertheless, the key motivation of this work (namely, the introduction of IR cutoff doublets in the power spectra) is adequately captured within our basic theoretical framework.

Next, we examined the two-point correlation function $C^{\rm EE}(\theta)$ for E-mode polarization, derived from {\it Planck} data, restricted to $\ell \leq 30$, directly comparable to the recent analyses carried out in \cite{Liu-Melia:2025}. Once again we find that introducing IR doublets in our analysis significantly reduces the discrepancy between theoretical predictions and observations, compared to the no-cutoff and single-cutoff scenarios (see Fig.~4).

Finally, we examined the B-mode polarization using the two-point correlation function $C^{\rm BB}(\theta)$, restricted to $\ell \leq 11$, as a potential tool to distinguish between different possible topologies of the early universe. In Figs.~5 and 6, we presented several representative plots for different choices of the tensor IR cutoff doublet and the tensor-to-scalar ratio, $r$, highlighting both the potential and the limitations of this approach, particularly at large angular scales where parity-odd preference would preferently manisfest.

Of course, new physics that violate parity can also be probed through other frameworks and observables \cite{Lue:1998mq,Feng:2004mq,Liu:2006uh,Saito:2007kt,Contaldi:2008yz}. Our approach, however, predicts specific signatures in the temperature anisotropies as well as in the E- and B-mode polarization correlations, which constitute independent types of measurements. Moreover, an eventual coincidence of odd-parity preference across all of these observables would therefore be unlikely to arise from a mere statistical fluctuation.

We conclude that forthcoming high-precision measurements of the CMB temperature and polarization from {\it LiteBIRD} \cite{LiteBIRD:2022cnt} could provide new insights into the tantalizing possibility of probing the topology of the early universe, as proposed in this work.

\section*{Acknowledgments}
I warmly thank Jingwei Liu and Verónica Sanz for illuminating discussions. I am also grateful to Jingwei Liu for sharing some plots prior to publication. This research was supported by 
the Spanish Agencia Estatal de Investigacion, under Grant PID2023-151418NB-I00 funded by MCIU/AEI/10.13039/501100011033/ FEDER, UE, and by GV under grant CIPROM/2022/36.

\end{document}